\documentclass[12pt,a4paper]{article}
\usepackage{latexsym}
\usepackage{amsfonts,amsbsy}

\def\Gammabol{{\stackrel{\circ}{\Gamma}}{}}

\def\Rbol{{\stackrel{\circ}{R}}{}}

\def\be{\begin{equation}}
\def\ee{\end{equation}}
\def\ba{\begin{eqnarray}}
\def\ea{\end{eqnarray}}
\begin{document}
\noindent
{\Large \bf Gravitation Without the Equivalence Principle}
\vskip 1.0cm
\noindent
{\bf R. Aldrovandi,\footnote{Instituto de F\'{\i}sica Te\'orica,
Universidade Estadual Paulista, Rua Pamplona 145,
01405-900 S\~ao Paulo, Brazil} J. G. Pereira$^1$ and K. H. Vu}$^1$
\vskip 2.0cm

\begin{abstract}
\noindent 
In the general relativistic description of gravitation, geometry replaces
the concept of force. This is possible because of the universal character of free
fall, and would break down in its absence. On the other hand, the teleparallel
version of general relativity is a gauge theory for the translation group and, as
such, describes the gravitational interaction by a force similar to the Lorentz
force of electromagnetism, a  non-universal interaction. Relying on this analogy it
is shown that, although the geometric description of general relativity necessarily
requires the existence of the equivalence principle, the teleparallel gauge approach
remains a consistent theory for gravitation in its absence.
\end{abstract}

\vskip 1.5cm
\section{Introduction}

Gravitation, like the other fundamental interactions of nature, can be described in terms
of a gauge theory \cite{PR}. In fact, the teleparallel equivalent of general relativity
\cite{tegra}, or teleparallel gravity for short \cite{obs,op}, corresponds to a gauge
theory for the translation group. In this approach, the gravitational interaction is
described by a force equation \cite{sp1} similar to the Lorentz force equation of
electrodynamics. On the other hand, due to the universality of free fall, it is also
possible to describe gravitation not as a {\em force}, but as a geometric {\em
deformation} of flat Minkowski spacetime. According to this point of view a gravitational
field produces a {\em curvature} in spacetime, and its action on (structureless)
particles is described by letting them follow the geodesics of the curved spacetime. This
is the approach of general relativity, in which geometry replaces the concept of
gravitational force, and the trajectories are determined, not by force equations, but by
geodesics.

Universality of free fall is then the reason for gravitation to present, in addition to
the teleparallel gauge approach, the equivalent geometric description of general
relativity. In fact, in order to attribute gravitation to curvature, it is essential that
gravitation be universal, or equivalently, that the {\em weak} equivalence principle,
which establishes the equality of {\em inertial} and {\em gravitational} masses, be true.
Only under these circumstances is it possible to assure that all particles of nature,
independently of their internal constitution, feel gravitation the same and, for a
given set of initial conditions, follow the same trajectory --- a geodesic of
the underlying Riemannian spacetime. 

Now, as is well known, the electromagnetic interaction is not universal, a property
consistent with the fact that there is no an electromagnetic equivalence principle. In
spite of this, Maxwell's theory, a gauge theory for the unitary group $U(1)$, is able to
consistently describe the electromagnetic interaction. Given the analogy between
electromagnetism and teleparallel gravity, the question then arises whether the gauge
approach of teleparallel gravity would also be able to describe the gravitational
interaction in the lack of universality, that is, in the absence of the weak equivalence
principle. The basic purpose of this paper is to give an answer to this question. It is
important to make it clear that, although there are many controversies related to the
equivalence principle \cite{synge,damour}, it is not our intention here to question its
validity, but simply verify whether the teleparallel description of gravitation requires or
not its existence. We begin by reviewing, in the next section, the basic concepts of the
teleparallel equivalent of general relativity.

\section{Fundamentals of Teleparallel Gravity}

The teleparallel equivalent of general relativity can be understood as a gauge theory of
the translation group. According to this theory, to each point of spacetime there is
attached a Minkowski tangent space, on which the translation (gauge) group acts. We use
the Greek alphabet $\mu, \nu, \rho, \dots = 0, 1, 2, 3$ to denote spacetime indices, and
the Latin alphabet $a, b, c, \dots = 0, 1, 2, 3$ to denote anholonomic indices related to
the tangent Minkowski spaces, whose metric is chosen to be $\eta_{a b} = {\rm diag} (+1,
-1, -1, -1)$.

The translational gauge potential $B^a{}_\mu$ is a 1-form assuming values in the Lie
algebra of the translation group, that is,
\be
B_\mu = B^a{}_\mu \, P_a,
\ee
with $P_a = \partial_a$ the generators of infinitesimal translations. This
potential appears as the nontrivial part of the tetrad field $h^{a}{}_{\mu}$,
which can then be written in the form
\be
h^a{}_\mu = \partial_\mu x^a + B^a{}_\mu.
\label{tetrada}
\ee
Notice that, whereas the tangent space indices are raised and lowered with the Minkowski
metric $\eta_{a b}$, the spacetime indices are raised and lowered with the spacetime
metric
\be
g_{\mu \nu} = \eta_{a b} \; h^a{}_\mu \; h^b{}_\nu.
\label{gmn}
\ee
Now, the above tetrad gives rise to the so called Weit\-zen\-b\"ock connection
\begin{equation}
\Gamma^{\rho}{}_{\mu\nu} = h_{a}{}^{\rho}\partial_{\nu}h^{a}{}_{\mu},
\label{carco}
\end{equation}
which introduces the distant parallelism on the four-dimensional spacetime manifold. It
is a connection that presents torsion, but no curvature. Its torsion,
\begin{equation}
T^{\rho}{}_{\mu\nu} = \Gamma^{\rho}{}_{\nu\mu} - 
\Gamma^{\rho}{}_{\mu\nu},
\label{tor}
\end{equation}
as we are going to see, turns out to be related to the translational gauge field strength.
The Weitzenb\"ock connection can be decomposed as
\begin{equation}
\Gamma^{\rho}{}_{\mu\nu} = \Gammabol^{\rho}{}_{\mu\nu} 
+ K^{\rho}{}_{\mu\nu},
\label{rela}
\end{equation}
where $\Gammabol^{\rho}{}_{\mu\nu}$ is the Christoffel connection constructed from the
spacetime metric $g_{\mu\nu}$, and
\begin{equation}
K^{\rho}{}_{\mu \nu} = \frac{1}{2} \left( 
T_{\mu}{}^{\rho}{}_{\nu} + T_{\nu}{}^{\rho}{}_{\mu} 
- T^{\rho}{}_{\mu \nu} \right)
\label{contorsion}
\end{equation}
is the contortion tensor. It is important to remark that we are considering curvature and
torsion as properties of a connection, not of spacetime \cite{livro}. Notice, for
example, that the Christoffel and the Weitzenb\"ock connections are defined on the same
spacetime metric manifold.

The Lagrangian of the teleparallel equivalent of general relativity is \cite{sp1}
\begin{equation}
{\mathcal L}_G = \frac{c^{4} h}{16\pi G} \, S^{\rho\mu\nu}\,T_{\rho\mu\nu} +
{\mathcal L}_M,
\label{gala}
\end{equation}
where $h = {\rm det}(h^{a}{}_{\mu})$, ${\mathcal L}_M$ is the Lagrangian of a source
field, and
\begin{equation}
S^{\rho\mu\nu} = - S^{\rho\nu\mu} = \frac{1}{2} 
\left[ K^{\mu\nu\rho} - g^{\rho\nu}\,T^{\sigma\mu}{}_{\sigma} 
+ g^{\rho\mu}\,T^{\sigma\nu}{}_{\sigma} \right]
\label{S}
\end{equation}
is a tensor written in terms of the Weitzenb\"ock connection only. Performing a variation
with respect to the gauge potential, we find the teleparallel version of the
gravitational field equation \cite{sp2},
\begin{equation}
\partial_\sigma(h S_\lambda{}^{\rho \sigma}) -
\frac{4 \pi G}{c^4} \, (h t_\lambda{}^\rho) =
\frac{4 \pi G}{c^4} \, (h {\mathcal T}_\lambda{}^\rho),
\label{eqs1}
\end{equation}
where
\begin{equation}
h \, t_\lambda{}^\rho = \frac{c^4 h}{4 \pi G} \, S_{\mu}{}^{\rho \nu}
\,\Gamma^\mu{}_{\nu\lambda} - \delta_\lambda{}^\rho \, {\mathcal L}_G
\label{emt}
\end{equation}
is the energy-momentum (pseudo) tensor of the gravitational field, and ${\mathcal
T}_\lambda{}^\rho = {\mathcal T}_a{}^\rho \, h^a{}_\lambda$ is the energy-momentum tensor
of the source field, with
\be
h \, {\mathcal T}_a{}^\rho = -
\frac{\delta {\mathcal L}_M}{\delta B^a{}_\rho} \equiv -
\frac{\delta {\mathcal L}_M}{\delta h^a{}_\rho}.
\ee
A solution of the gravitational field
equation (\ref{eqs1}) is an explicit form of the gravitational gauge potential
$B^a{}_\mu$.

When the weak equivalence principle is assumed to be true, teleparallel gravity turns
out to be equivalent to general relativity. In fact, up to a divergence, the Lagrangian
(\ref{gala}) is equivalent to the Einstein-Hilbert Lagrangian of general relativity,
\be
{\mathcal L}_G = \frac{c^{4} h}{16\pi G} \; \Rbol,
\ee
with $\Rbol$ the scalar curvature of the Christoffel connection. Accordingly, the
teleparallel field equation (\ref{eqs1}) is found to coincide with Einstein's equation
\be
\Rbol_\lambda{}^\rho - \frac{1}{2} \delta_\lambda{}^\rho \Rbol =
\frac{8 \pi G}{c^4} \, {\mathcal T}_\lambda{}^\rho,
\label{eeq}
\ee
where $\Rbol_\lambda{}^\rho$ is the Ricci curvature of the Christoffel connection. Lets
us then see what happens when the weak equivalence principle is assumed not to be true.

\section{Teleparallel Equation of Motion}

To begin with, let us consider, in the context of teleparallel gravity, the motion of a
spinless particle in a gravitational field $B^{a}{}_{\mu}$, supposing however that the
gravitational mass $m_g$ and the inertial mass $m_i$ do not coincide. Analogously to the
electromagnetic case \cite{landau}, the action integral is written in the form
\be
S = \int_{a}^{b} \left[ - m_i \, c \, d\sigma -
m_g \, c \, B^{a}{}_{\mu} \, u_{a} \, dx^{\mu} \right],
\label{acao1}
\ee
where $d\sigma = (\eta_{a b} dx^a dx^b)^{1/2}$ is the Minkowski tangent-space invariant
interval, and $u^a$ is the particle four-velocity seen from the tetrad frame,
necessarily anholonomic when expressed in terms of the {\it spacetime} line element $ds$
\cite{ABP03}. It should be noticed, however, that in terms of the {\it tangent-space} line
element $d
\sigma$, it is holonomic, that is (see Appendix)
\be
u^a = \frac{d x^a}{d \sigma}.
\label{native}
\ee
The first term of the action (\ref{acao1}) represents the action of a free particle, and
the second the coupling of the particle with the gravitational field. Notice that the
separation of the action in these two terms is possible only in a gauge theory, like
teleparallel gravity, being not possible in general relativity.

Variation of the action (\ref{acao1}) yields
\ba
\delta S = \int_{a}^{b} m_i c \Big[ \Big( \partial_\mu x^a +
\frac{m_g}{m_i} \; B^a{}_\mu \Big) \frac{d u_a}{d s}
&-& \frac{m_g}{m_i} \; (\partial_\mu B^a{}_\rho -
\partial_\rho B^a{}_\mu ) u_a \, u^\rho \nonumber \\
&-& \frac{m_g}{m_i} \; B^a{}_\rho \; \partial_\mu u_a \, u^\rho \Big]
\delta x^\mu \, ds,
\label{delta1}
\ea
where
\be
u^\mu = \frac{d x^\mu}{ds} \equiv h^\mu{}_a \, u^a
\label{ust}
\ee
is the particle four-velocity, with $ds$ = $(g_{\mu \nu} dx^\mu dx^\nu)^{1/2}$ the
Riemannian spacetime invariant interval. As $B^a{}_\mu$ is an Abelian gauge potential,
\be
\partial_\mu B^a{}_\rho -
\partial_\rho B^a{}_\mu \equiv F^a{}_{\mu \rho}
\label{tfs}
\ee
will be the corresponding gravitational field strength. Using Eqs.\ (\ref{tetrada}) and
(\ref{carco}), we see that $F^a{}_{\mu \rho}$ is nothing but the torsion of the
Weitzenb\"ock connection with the first index written in the tetrad basis:
\be
F^a{}_{\mu \rho} = h^a{}_\lambda \, T^\lambda{}_{\mu \rho}.
\label{fstor}
\ee
Finally, the last term in Eq.~(\ref{delta1}) does not contribute to the equation of
motion. Indeed, after substituting $B^a{}_\rho$ = $h^a{}_\rho - \partial_\rho x^a$ and
using Eq.~(\ref{native}), it becomes proportional to
\be
\int_{a}^{b} \left[ \left(1 - \frac{d \sigma}{d s} \right) u^a \; \partial_\mu u_a
\right] \delta x^\mu \, ds,
\ee
which vanishes because $u^a u_a = 1$. We are then left with
\be
\delta S = \int_{a}^{b} m_i c \Big[ \Big( \partial_\mu x^a +
\frac{m_g}{m_i} \; B^a{}_\mu \Big) \frac{d u_a}{d s}
- \frac{m_g}{m_i} \; F^a{}_{\mu \rho} \; u_a \, u^\rho \Big]
\delta x^\mu \, ds.
\label{delta3}
\ee
From the invariance of the action, and taking into account the arbitrariness of $\delta
x^\mu$, we get
\be
\left( \partial_\mu x^a +
\frac{m_g}{m_i} \; B^a{}_\mu \right) \frac{d u_a}{d s} =
\frac{m_g}{m_i} \; F^a{}_{\mu \rho} \; u_a \, u^\rho.
\label{eqmot2}
\ee
This is the force equation governing the motion of the particle, in which the
teleparallel field strength $F^a{}_{\mu \rho}$ plays the role of gravitational force.
Similarly to the electromagnetic Lorentz force equation, which depends on the relation
$e/m_i$, with $e$ the electric charge of the particle, the gravitational force equation
depends explicitly on the relation ${m_g}/{m_i}$ of the particle. When $m_g=m_i$, it is
easily seen to coincide with the geodesic equation of general relativity.

The above results show that, even in the absence of the weak equivalence principle,
teleparallel gravity is able to describe the motion of a particle with $m_g \neq m_i$.
The crucial point is to observe that, although the equation of motion depends explicitly
on the relation $m_i/m_g$ of the particle, neither $B^a{}_\mu$ nor $F^a{}_{\rho \mu}$
depends on this relation. This means essentially that the teleparallel field equation
(\ref{eqs1}) can be consistently solved for the gravitational potential $B^a{}_\mu$,
which can then be used to write down the equation of motion (\ref{eqmot2}), independently
of the validity or not of the weak equivalence principle. The gauge potential $B^a{}_\mu$,
therefore, may be considered as the most fundamental field representing gravitation. As we
are going to see next, this is not the case of general relativity, in which the
gravitational field necessarily depends on the relation $m_i/m_g$ of the particle,
rendering thus the theory inconsistent.

\section{Relation with General Relativity}

By using the relation (\ref{fstor}), as well as the identity
\be
T^\lambda{}_{\mu \rho} \, u_\lambda \, u^\rho = - K^\lambda{}_{\mu \rho}
\, u_\lambda \, u^\rho,
\ee
the force equation (\ref{eqmot2}) can be rewritten in the form
\be
\frac{d u_\mu}{ds} - \Gammabol^\lambda{}_{\mu \rho} \, u_\lambda \, u^\rho =
\left(\frac{m_g - m_i}{m_g} \right) \partial_\mu x^a \, \frac{d u_a}{d s},
\label{eqmot6}
\ee
where use has been made also of the relation (\ref{rela}). Notice that the violation of
the weak equivalence principle produces a deviation from the geodesic motion, which is
proportional to the difference between the gravitational and inertial masses. Notice
furthermore that, due to the assumed non-universality of free fall, there is no a local
coordinate system in which the gravitational effects are absent.

Now, as already said, when the weak equivalence principle is assumed to be true, the
teleparallel field equation (\ref{eqs1}) is equivalent to the Einstein equation
(\ref{eeq}). Accordingly, when $m_g = m_i$, the equation of motion (\ref{eqmot2}) reduces
to the geodesic equation of general relativity, as can be seen from its equivalent form
(\ref{eqmot6}). However, in the absence of the weak equivalence principle, it is not a
geodesic equation. This means that the equation of motion (\ref{eqmot2}) does not comply
with the geometric description of general relativity, according to which all trajectories
must be given by genuine geodesic equations. In order to comply with the foundations of
general relativity, it is necessary to incorporate the particle properties into the
geometry. This can be achieved by assuming, instead of the tetrad (\ref{tetrada}) of
teleparallel gravity, the new tetrad
\be
\bar{h}^a{}_\mu = \partial_\mu x^a + \frac{m_g}{m_i} \; B^a{}_\mu,
\label{tetrada2}
\ee
which takes into account the characteristic $m_g/m_i$ of the particle under consideration.
This tetrad defines a new spacetime metric tensor
\be
\bar{g}_{\mu \nu} = \eta_{a b} \; \bar{h}^a{}_\mu \; \bar{h}^b{}_\nu,
\label{gmn2}
\ee
in terms of which the corresponding spacetime invariant interval is
\be
d\bar{s}^2 = \bar{g}_{\mu \nu} \, dx^\mu dx^\nu.
\ee
By noticing that in this case the relation between the gravitational field strength and
torsion turns out to be
\be
\frac{m_g}{m_i} \; F^a{}_{\mu \rho} = \bar{h}^a{}_\lambda \, \bar{T}^\lambda{}_{\mu \rho},
\label{fstor2}
\ee
it is an easy task to verify that, for a fixed relation $m_g/m_i$, the equation of motion
(\ref{eqmot2}) is equivalent to the true geodesic equation
\be
\frac{d \bar{u}_\mu}{d\bar{s}} - {\bar{\Gamma}}{}^\lambda{}_{\mu \rho}
\, \bar{u}_\lambda \, \bar{u}^\rho = 0,
\label{eqmot7}
\ee
where $\bar{u}_\mu \equiv d x_\mu/d \bar{s} = \bar{h}^a{}_\mu u_a$, and
$\bar{\Gamma}{}^{\rho}{}_{\mu\nu}$ is the Christoffel connection of the metric
$\bar{g}_{\mu \nu}$. Notice that this equation can also be obtained from the action
integral
\be
\bar{S} = - m_i \ c \int_a^b d\bar{s},
\ee
which is the usual form of the action in the context of general relativity.

However, the price for imposing a geodesic equation of motion to describe a non-universal
interaction is that the gravitational theory becomes inconsistent. In fact, the solution
of the corresponding Einstein's field equation
\be
\bar{R}_{\mu \nu} - \frac{1}{2} \, \bar{g}_{\mu \nu} \bar{R} =
\frac{8 \pi G}{c^4} \, \bar{\mathcal T}_{\mu \nu},
\label{e2}
\ee
which is not equivalent to any teleparallel field equation, would in this case depend on
the relation $m_g/m_i$ of the test particle, which renders the theory inconsistent in the
sense that test particles with different relations $m_g/m_i$ would require connections
with different curvatures to keep all equations of motion given by geodesics. Of course,
as a true field, the gravitational field cannot depend on any test particle properties.

\section{Final Remarks}

In Einstein's general relativity, which is a theory fundamentally based on the
universality of free fall, or equivalently, on the weak equivalence principle, geometry
replaces the concept of force in the description of the gravitational interaction. In spite
of the fact that, at least at the classical level, it has passed all experimental tests
\cite{exp}, a possible violation of the weak equivalence principle, among other observable
consequences, would lead to the non-universality of free fall, and consequently to the ruin
of the general relativity description of gravitation. We notice in passing that the absence
of an electromagnetic equivalence principle is the reason why there is no a geometric
description, in the sense of general relativity, for the electromagnetic interaction.

On the other hand, as a gauge theory for the translation group, the teleparallel
equivalent of general relativity does not describe the gravitational interaction
through a geometrization of spacetime, but as a gravitational force quite analogous to
the Lorentz force equation of electrodynamics. In the same way Maxwell's gauge theory is
able to describe the non-universal electromagnetic interaction, we have shown that
teleparallel gravity is also able to describe the gravitational interaction in the absence
of universality, remaining in this way a consistent theory for gravitation. In spite of the
equivalence between the geometric description of general relativity and the gauge
description of teleparallel gravity when the weak equivalence principle is assumed to hold
\cite{always}, the latter can be considered as a more fundamental theory in the sense that
it has no need of the principle to describe the gravitational interaction. Notice in this
connection that the equivalence principle is frequently said to preclude the definition of
a local energy-momentum density for the gravitational field \cite{gravitation}. Although
this is a true assertion in the context of general relativity, it has already been
demonstrated that, in the gauge context of teleparallel gravity, a tensorial expression
for the gravitational energy-momentum density is possible \cite{sp2}, which shows the
consistency of our results. 

On the strength of our results, we can say that the fundamental field describing
gravitation is neither the tetrad nor the metric, but the translational gauge potential
$B^a{}_\mu$. This point may have important consequences for both classical and quantum
gravity. For example, gravitational waves should be interpreted as $B$ waves and
not as metric waves as this is not a fundamental, but a derived quantity. For the same
reason, the quantization of the gravitational field should be carried out on $B^a{}_\mu$
and not on the tetrad or on the metric fields. Another important consequence refers to a
fundamental problem of quantum gravity, namely, the conceptual difficulty of reconciling
{\it local} general relativity with {\it non-local} quantum mechanics, or equivalently, of
reconciling the local character of the equivalence principle with the non-local character
of the uncertainty principle \cite{qep}. As far as teleparallel gravity can be formulated
independently of any equivalence principle, the quantization of the gravitational field may
possibly appear much more consistent if considered in the teleparallel approach.

\section*{Acknowledgments}

The authors would like to thank FAPESP-Brazil, CNPq-Brazil, and CAPES-Brazil for
financial support.

\begin{appendix}

\section*{Appendix}
According to our notation, $d\sigma = (\eta_{a b} dx^a dx^b)^{1/2}$ represents the
Minkowski tangent-space invariant interval, and $ds$ = $(g_{\mu \nu} dx^\mu
dx^\nu)^{1/2}$ the spacetime invariant interval. Now, instead of working with quadratic
intervals, it is far more convenient to introduce the Dirac matrices $\gamma^a =
h^a{}_\mu \gamma^\mu$ and, similarly to the Dirac equation, work with the {\it linear}
matrix form of the intervals. In terms of the $\gamma$ matrices, the spacetime
and the tangent-space matrix invariant intervals are respectively \cite{fock}
\be
\hat{ds} = \gamma_\mu \, dx^\mu \equiv \gamma_a \, h^a{}_\mu \, d x^\mu
\ee
and
\be
\hat{d \sigma} = \gamma_a \, d x^a .
\label{dsm}
\ee
In the language of differential forms,
\[
\gamma_a = \hat{d \sigma} \left(\frac{\partial}{\partial x^a}
\right) = \frac{\hat{d \sigma}}{d x^a}
\]
and consequently
\[
\hat{d s} \left(\frac{\partial}{\partial x^\mu} \right) = \frac{\hat{d s}}{d x^\mu} =
\gamma_a \, h^a{}_\mu  = h^a{}_\mu \frac{\hat{d \sigma}}{d x^a}.
\]
Using the relation ${\partial}/{\partial x^\mu} = ({\partial x^a}/{\partial x^\mu})
{\partial}/{\partial x^a}$, we get
\be
\hat{d s} \, \partial_\mu x^a = \hat{d \sigma} \, h^a{}_\mu.
\label{mare}
\ee

We have now to return from the matrix to the usual form of the interval. This can be
achieved by taking the determinant on both sides of Eq. (\ref{mare}). By
using that
\be
\det(\hat{ds}) = (ds)^4 \quad {\rm and} \quad \det(\hat{d\sigma}) = (d\sigma)^4,
\ee
we obtain immediately
\be
ds \, \partial_\mu x^a = d \sigma \, h^a{}_\mu.
\ee
Equivalently, we can write
\be
h^a{}_\mu = \frac{d s}{d \sigma} \, \frac{\partial x^a}{\partial x^\mu}.
\label{spetetra}
\ee
The inverse tetrad is consequently
\be
h^\mu{}_a = \frac{d \sigma}{ds} \, \frac{\partial x^\mu}{\partial x^a}.
\ee
Of course, these expressions are valid only along the trajectory of the particle. Notice
in addition that, in these forms, the tetrads represent a measure of how much $ds$ and
$d \sigma$ differ from each other. In the absence of gravitation, $ds = d \sigma$, and the
tetrad becomes trivial.

On the other hand, we know that (see Eq. (\ref{ust}))
\be
u^a = h^a{}_\mu \, u^\mu \equiv h^a{}_\mu \, \frac{dx^\mu}{ds}.
\ee
Substituting (\ref{spetetra}), we obtain that, along the trajectory,
\be
u^a = \frac{\partial x^a}{\partial x^\mu} \, \frac{dx^\mu}{ds} \,
\frac{d s}{d \sigma} \equiv \frac{dx^a}{d\sigma},
\ee
which is expression (\ref{native}) of the particle four-velocity.
\end{appendix} 


\end{document}